\documentclass[prb,aps,amssymb,twocolumn]{revtex4}
\usepackage{graphicx}% Include figure files

\begin{document}

\title{The quantum spin Hall effect and topological insulators}

\author{Xiao-Liang Qi and Shou-Cheng Zhang}

 \affiliation{Department of Physics, McCullough
Building, Stanford University, Stanford, CA 94305-4045\\
Stanford Institute for Materials and Energy Sciences, SLAC National
Accelerator Laboratory, 2575 Sand Hill Road, Menlo Park, CA 94025,
USA.}

\date{\today}
\begin{abstract}
In topological insulators, spin-orbit coupling and time-reversal
symmetry combine to form a novel state of matter predicted to have
exotic physical properties.
\end{abstract}
\maketitle

In the quantum world, atoms and their electrons can form many
different states of matter, such as crystalline solids, magnets, and
superconductors. Those different states can be classified by the
symmetries they spontaneously break---translational, rotational, and
gauge symmetries, respectively, for the examples above. Before 1980
all states of matter in condensed matter systems could be classified
by the principle of broken symmetry. The quantum Hall (QH) state,
discovered in 1980,\cite{r1} provided the first example of a quantum
state that has no spontaneously broken symmetry. Its behavior
depends only on its topology and not on its specific geometry; it
was topologically distinct from all previously known states of
matter.

Recently, a new class of topological states has emerged, called
quantum spin Hall (QSH) states or topological insulators (see
PHYSICS TODAY, January 2008, page 19). Topologically distinct from
all other known states of matter, including QH states, QSH states
have been theoretically predicted and experimentally observed in
mercury telluride quantum wells,\cite{r2,r3} in bismuth antimony
alloys,\cite{r4,r5} and in ${\rm Bi_2Se_3}$ and ${\rm Bi_2Te_3}$
bulk crystals.\cite{r6,r7,r8} QSH systems are insulating in the
bulk---they have an energy gap separating the valence and conduction
bands---but on the boundary they have gapless edge or surface states
that are topologically protected and immune to impurities or
geometric perturbations.\cite{r9,r10,r11,r12} Inside such a
topological insulator, Maxwell¡¯s laws of electromagnetism are
dramatically altered by an additional topological term with a
precisely quantized coefficient,\cite{r12} which gives rise to
remarkable physical effects. Whereas the QSH state shares many
similarities with the QH state, it differs in important ways. In
particular, QH states require an external magnetic field, which
breaks time reversal (TR) symmetry; QSH states, in contrast, are TR
invariant and do not require an applied field.

\section{From quantum Hall to quantum spin Hall}

In a one-dimensional world, there are two basic motions: forward and
backward. Random scattering can cause them to mix, which leads to
resistance. Just as we have learned from basic traffic control, it
would be much better if we could spatially separate the counterflow
directions into two separate lanes, so that random collisions could
be easily avoided. That simple traffic control mechanism turns out
to be the essence of the QH effect.\cite{r1}

The QH effect occurs when a strong magnetic field is applied to a 2D
gas of electrons in a semiconductor. At low temperature and high
magnetic field, electrons travel only along the edge of the
semiconductor, and the two counterflows of electrons are spatially
separated into different ``lanes" located at the sample's top and
bottom edges. Compared with a 1D system with electrons propagating
in both directions, the top edge of a QH bar contains only half the
degrees of freedom. That unique spatial separation is illustrated in
figure 1a by the symbolic equation ``2 = 1 [forward mover] + 1
[backward mover]" and is the key reason why the QH effect is
topologically robust. When an edge-state electron encounters an
impurity, it simply takes a detour and still keeps going in the same
direction (figure 1), as there is no way for it to turn back. Such a
dissipationless transport mechanism could be extremely useful for
semiconductor devices. Unfortunately, the requirement of a large
magnetic field severely limits the application potential of the QH
effect.

Can we get rid of the magnetic field and still separate the traffic
lanes for the electrons? In a real 1D system, forward and
backward-moving channels for both spin-up and spin-down electrons
give rise to four channels, as shown in figure 1b. The traffic lanes
for the electrons can be split in a TR-invariant fashion, without
any magnetic field, as illustrated in the figure by the symbolic
equation ``4 = 2 + 2." We can leave the spin-up forward mover and
the spin-down backward mover on the top edge and move the other two
channels to the bottom edge. A system with such edge states is said
to be in a QSH state, because it has a net transport of spin forward
along the top edge and backward along the bottom edge, just like the
separated transport of charge in the QH state. Charles Kane and
Eugene Mele from the University of Pennsylvania,\cite{r9} and Andrei
Bernevig and one of us (Zhang)\cite{r10} from Stanford University,
independently proposed in 2005 and 2006 that such a separation, and
thus the QSH state, can in principle be realized in certain
theoretical models with spin-orbit coupling. (The fractional QSH
state was also predicted,\cite{r10} though it has yet to be
experimentally observed.)

Although a QSH edge consists of both backward and forward movers,
backscattering by nonmagnetic impurities is forbidden. To understand
that effect, we start with an analogy from daily life. Most
eyeglasses and camera lenses have a so-called antireflection
coating. As shown in figure 2a, reflected light from the top and the
bottom surfaces interfere with each other destructively, leading to
zero net reflection and thus perfect transmission. However, such an
effect is not robust, as it depends on the matching between the
optical wavelength and the thickness of the coating.

Just like the reflection of a photon by a surface, an electron can
be reflected by an impurity, and different reflection paths also
interfere with each other. As shown in figure 2b, an electron in a
QSH edge state can take either a clockwise or a counterclockwise
turn around the impurity, and during that turn the spin rotates by
an angle of $\pi$ or $-\pi$ to the opposite direction. Consequently,
the two paths, related by TR symmetry, differ by a full
$\pi-(-\pi)=2\pi$ rotation of the electron spin. A profound and yet
deeply mysterious principle of quantum mechanics states that the
wavefunction of a spin-$1/2$ particle obtains a negative sign upon a
full $2\pi$ rotation. Thus the two backscattering paths always
interfere destructively, which leads to perfect transmission. If the
impurity carries a magnetic moment, the TR symmetry is broken and
the two reflected waves no longer interfere destructively. In that
sense the robustness of the QSH edge state is protected by the TR
symmetry.

The physical picture above applies only to the case of single pairs
of QSH edge states. If there are two forward movers and two backward
movers in the system---as, for example, the unseparated 1D system
shown in figure 1b---then an electron can be scattered from a
forward- to a backward-moving channel without reversing its spin and
without the perfect destructive interference, and thus there is
dissipation. Consequently, for the QSH state to be robust, the edge
states must consist of an odd number of forward movers and an odd
number of backward movers. That even¨Codd effect, characterized by a
so-called ${\rm Z_2}$ topological quantum number, is at the heart of
the QSH state\cite{r9,r13} and is why a QSH insulator is also
synonymously referred to as a topological insulator.

\begin{figure}[htpb]
    \begin{center}
        \includegraphics[width=3.5in]{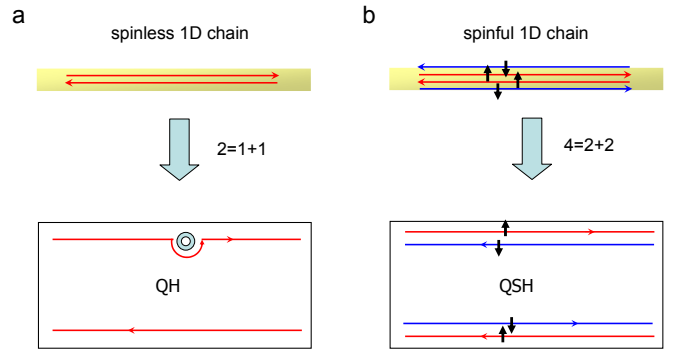}
    \end{center}
    \caption{Spatial separation is at the heart of both the quantum Hall (QH) and
the quantum spin Hall (QSH) effects. (a) A spinless one-dimensional
system has both a forward and a backward mover. Those two basic
degrees of freedom are spatially separated in a QH bar, as
illustrated by the symbolic equation ``2 = 1 + 1." The upper edge
contains only a forward mover and the lower edge has only a backward
mover. The states are robust: They will go around an impurity
without scattering. (b) A spinful 1D system has four basic channels,
which are spatially separated in a QSH bar: The upper edge contains
a forward mover with up spin and a backward mover with down spin,
and conversely for the lower edge. That separation is illustrated by
the symbolic equation ``4 = 2 + 2."}
    %\caption{{\bf a}. Picture of a Hall bar ({\bf citation?}). {\bf
%    b}. Typical curve of Hall resistance $R_{xy}$ versus magnetic
%    field in an integer quantum Hall system. $i=2,3,4...$ stand for the integer in the quantized Hall resistance $R_{xy}=h/ie^2$, which
%    corresponds to the number of filled Landau levels. {\bf c}. Comparison between the edge states of
%    integer quantum Hall system (upper panel) and quantum spin Hall system (lower panel).
%    The edge states of quantum Hall system are ``chiral", meaning that they only propagate towards one direction on the edge; whereas the edge
%    states of quantum spin Hall system are ``helical", meaning that the states with spin up only propagate towards one direction while those with spin down only propagate
%    towards the opposite direction.}
    \label{fig1}
\end{figure}

\section{Two dimensional topological insulators}

Looking at figure 1b, we see that the QSH effect requires the
counterpropagation of opposite spin states. Such a coupling between
the spin and the orbital motion is a relativistic effect most
pronounced in heavy elements. Although all materials have spin-orbit
coupling, only a few of them turn out to be topological insulators.
In 2006 Bernevig, Taylor Hughes, and Zhang proposed a general
mechanism for finding topological insulators\cite{r2} and predicted
in particular that mercury telluride quantum wells---nanoscopic
layers sandwiched between other materials---are topological
insulators beyond a critical thickness $d_c$. The general mechanism
is band inversion, in which the usual ordering of the conduction
band and valence band is inverted by spin-orbit
coupling.\cite{r2,r4}

In most common semiconductors, the conduction band is formed from
electrons in $s$ orbitals and the valence band is formed from
electrons in $p$ orbitals. In certain heavy elements such as Hg and
Te, however the spin-orbit coupling is so large that the p-orbital
band is pushed above the s-orbital band---that is, the bands are
inverted. Mercury telluride quantum wells can be prepared by
sandwiching the material between cadmium telluride, which has a
similar lattice constant but much weaker spin-orbit coupling.
Therefore, increasing the thickness $d$ of the HgTe layer increases
the strength of the spin-orbit coupling for the entire quantum well.
For a thin quantum well, as shown in the left column of figure 3a,
the CdTe has the dominant effect and the bands have a normal
ordering: The s-like conduction subband E1 is located above the
p-like valence subband H1. In a thick quantum well, as shown in the
right column, the opposite ordering occurs due to increased
thickness d of the HgTe layer. The critical thickness $d_c$ for band
inversion is predicted to be around $6.5$ nm.

The QSH state in HgTe can be described by a simple model for the E1
and H1 subbands\cite{r2} (see the appendix). Explicit solution of
that model gives one pair of edge states for $d
> d_c$ in the inverted regime and no edge states in the $d < d_c$, as
shown in figure 3b. The pair of edge states carry opposite spins and
disperse all the way from valence band to conduction band. The
crossing of the dispersion curves is required by TR symmetry and
cannot be removed---it is one of the topological signatures of a QSH
insulator.

Less than one year after the theoretical prediction, a team at the
University of W$\ddot{u}$rzburg led by Laurens Molenkamp observed
the QSH effect in HgTe quantum wells grown by molecular-beam
epitaxy.\cite{r3} The edge states provide a direct way to
experimentally distinguish the QSH insulator from the trivial
insulator. The two edge states of the QSH insulator act as two
conducting 1D channels, which each contribute one quantum of
conductance, $e^2/h$. That perfect transmission is possible because
of the principle of antireflection explained earlier. In contrast, a
trivial insulator phase is ``really" insulating, with vanishing
conductance. Such a sharp conductance difference between thin and
thick quantum wells was observed experimentally, as shown in figure
3c.

\begin{figure}[htpb]
    \begin{center}
        \includegraphics[width=3in]{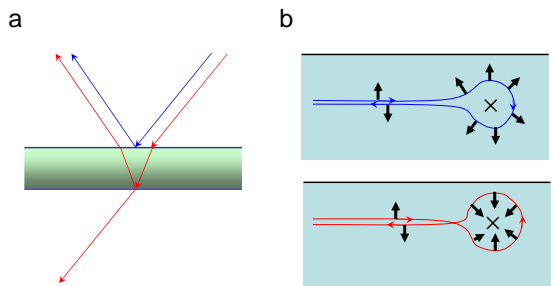}
    \end{center}
    \caption{(a) On a lens with antireflection
coating, light waves reflected by the top (blue line) and the bottom
(red line) surfaces interfere destructively, which leads to
suppressed reflection. (b) A quantum spin Hall edge state can be
scattered in two directions by a nonmagnetic impurity. Going
clockwise along the blue curve, the spin rotates by $\pi$;
counterclockwise along the red curve, by $-\pi$. A quantum
mechanical phase factor of $-1$ associated with that difference of
$2\pi$ leads to destructive interference of the two paths---the
backscattering of electrons is suppressed in a way similar to that
of photons off the antireflection coating.}
    \label{fig2}
\end{figure}

\section{From two to three dimensions}

\begin{figure}[htpb]
    \begin{center}
        \includegraphics[width=3in]{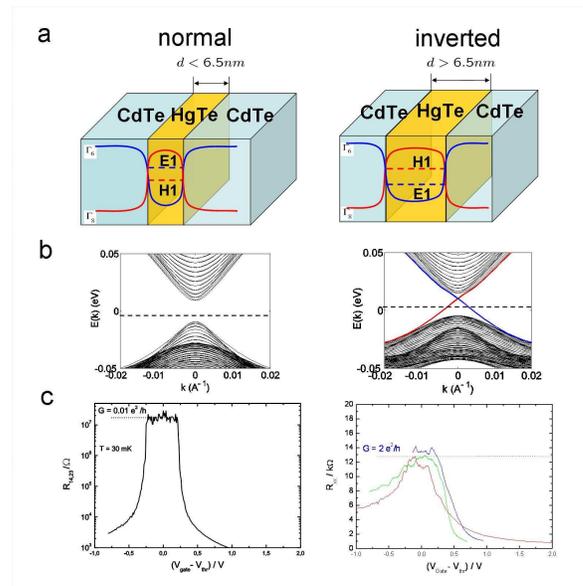}
    \end{center}
    \caption{Mercury telluride quantum wells are two-dimensional topological
insulators. (a) The behavior of a mercury telluride-cadmium
telluride quantum well depends on the thickness $d$ of the HgTe
layer. Here the blue curve shows the potential-energy well
experienced by electrons in the conduction band; the red curve is
the barrier for holes in the valence band. Electrons and holes are
trapped laterally by those potentials but are free in the other two
dimensions. For quantum wells thinner than a critical thickness
$d_c\simeq 6.5 {\rm nm}$, the energy of the lowest energy conduction
subband, labeled E1, is higher than that of the highest-energy
valence band, labeled H1. But for $d>d_c$, those electron and hole
bands are inverted. (b) The energy spectra of the quantum wells. The
thin quantum well has an insulating energy gap, but inside the gap
in the thick quantum well are edge states, shown by red and blue
lines. (c) Experimentally measured resistance of thin and thick
quantum wells, plotted against the voltage applied to a gate
electrode to change the chemical potential.\cite{r3} The thin
quantum well has a nearly infinite resistance within the gap,
whereas the thick quantum well has a quantized resistance plateau at
$R = h/2e^2$, due to the perfectly conducting edge states. Moreover,
the resistance plateau is the same for samples with different
widths, from 0.5 $\mu$m (red) to 1.0 $\mu$m (blue), proof that only
the edges are conducting.}
    \label{fig3}
\end{figure}

From figure 3b we see that the 2D topological insulator has a pair
of 1D edge states crossing at momentum $k = 0$. Near the crossing
point, the dispersion of the states is linear. That's exactly the
dispersion one gets in quantum field theory from the Dirac equation
for a massless relativistic fermion in 1D, and thus that equation
can be used to describe the QSH edge state. Such a picture can be
simply generalized to a 3D topological insulator, for which the
surface state consists of a single 2D massless Dirac fermion and the
dispersion forms a so-called Dirac cone, as illustrated in figure 4.
Similar to the 2D case, the crossing point---the tip of the
cone---is located at a TR-invariant point, such as at $k = 0$, and
the degeneracy is protected by TR symmetry.

Liang Fu and Kane predicted\cite{r4} that the alloy ${\rm
Bi_{1-x}Sb_x}$ would be a 3D topological insulator in a special
range of $x$, and with angle-resolved photoemission spectroscopy
(ARPES) Zahid Hasan and coworkers at Princeton University observed
the topological surface states in that system.\cite{r5} However, the
surface states and the underlying mechanism turn out to be extremely
complex. In collaboration with Zhong Fang¡¯s group at the Chinese
Academy of Sciences, the two of us predicted that ${\rm Bi_2Te_3}$,
${\rm Bi_2Se_3}$, and ${\rm Sb_2Te_3}$, all with the layered
structure in figure 4a, are 3D topological insulators, whereas a
related material, ${\rm Sb_2Se_3}$, is not.\cite{r6}

As in HgTe, the nontrivial topology of the ${\rm Bi_2Te_3}$ family
is due to band inversion between two orbitals with opposite parity,
driven by the strong spin-orbit coupling of Bi and Te. Due to such
similarity, that family of 3D topological insulators can be
described by a 3D version of the HgTe model (see the appendix).
First-principle calculations show that the materials have a single
Dirac cone on the surface. The spin of the surface state lies in the
surface plane and is always perpendicular to the momentum, as shown
in figure 4b.

Known to be excellent thermoelectric materials, ${\rm Bi_2Te_3}$ and
${\rm Bi_2Se_3}$ have been investigated independently, particularly
at Princeton, where Hasan¡¯s group observed in ARPES experiments the
single Dirac-cone surface state of ${\rm Bi_2Se_3}$ samples prepared
by Robert Cava and coworkers.\cite{r7} Furthermore, the group's
spin-resolved measurements showed that the electron spin indeed lies
in the plane of the surface and is always perpendicular to the
momentum, in agreement with theory. However, the experiments also
observed bulk carriers coexisting with the topological surface
states. A pure topological insulator phase without bulk carriers was
first observed in ${\rm Bi_2Te_3}$ by Yulin Chen and Zhi-xun Shen's
group at Stanford in ARPES experiments on material prepared by Ian
Fisher and colleagues.\cite{r8} As shown in figure 4c, the observed
surface states indeed disperse linearly, crossing at the point with
zero momentum. By mapping all of momentum space, the ARPES
experiments show convincingly that the surface states of ${\rm
Bi_2Te_3}$ and ${\rm Bi_2Se_3}$ consist of a single Dirac cone. Such
a state is impossible to construct in a purely 2D system. For
example, a 2D graphene sheet has four Dirac cones (see the article
by Andrey Geim and Allan MacDonald, PHYSICS TODAY, August 2007, page
35). The 2D HgTe quantum well at the crossover point $d=d_c$ has two
Dirac cones. In a sense, the cones are spatially separated, with one
placed on the top surface and the other on the bottom surface,
similar to spatial decomposition illustrated in figure 1 for the 1D
surfaces of a 2D system. (Particle physicists have been using
similar ideas---treating a 3D lattice as the surface of a 4D
lattice---in numerical simulation of fermions so as to avoid getting
an unwanted doubling of neutrinos.) As we discuss below, a single
Dirac cone on the surface directly leads to novel topological
properties.\cite{r11,r12}

\begin{figure}[htpb]
    \begin{center}
        \includegraphics[width=3in]{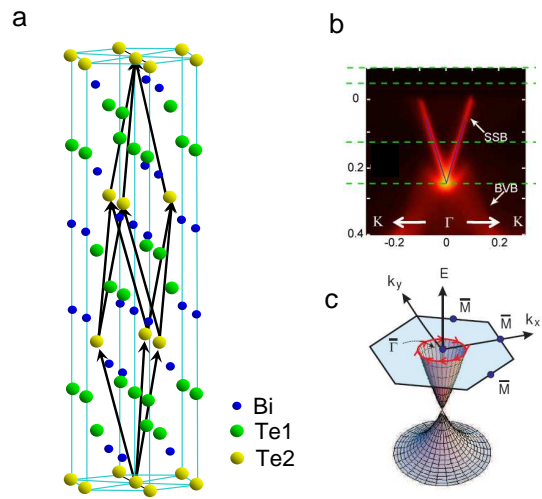}
    \end{center}
    \caption{In three-dimensional topological
insulators, the linearly dispersing edge states of figure 3b become
surface states described by a so-called Dirac cone. (a) The crystal
structure of the 3D topological insulator ${\rm Bi_2Te_3}$ consists
of stacked quasi-2D layers of Te-Bi-Te-Bi-Te. The arrows indicate
the lattice basis vectors. The surface state is predicted to consist
of a single Dirac cone.\cite{r6} (b) Angle-resolved photoemission
spectroscopy maps the energy states in momentum space. Spin
dependent ARPES of the related compound ${\rm Bi_2Se_3}$ reveals
that the spins (red) of the surface states lie in the surface plane
and are perpendicular to the momentum.\cite{r7} (c) This ARPES plot
of energy versus wavenumber in ${\rm Bi_2Te_3}$ shows the linearly
dispersing surface-state band (SSB) above the bulk valence band
(BVB). The top two dashed green lines denote the bulk insulating
gap; the bottom line marks the point of the Dirac cone.\cite{r8}}
    \label{fig4}
\end{figure}

\section{Topological classification of insulators}

Mathematicians group geometric objects into broad topological
classes. Objects with different shapes, such as a donut and a coffee
cup, can be smoothly deformed into each other and can therefore be
grouped into the same topological class. Mathematicians also
developed the concept of a topological invariant that uniquely
defines the topological class. Topological materials in general, and
topological insulators in particular, can be defined by physically
measurable topological invariants in topological field theories.

We can first divide insulators into two broad classes, according to
the presence or absence of TR symmetry. The QH state is a
topological insulator state that breaks TR symmetry. David Thouless
and coworkers showed that the physically measured integer QH
conductance is given by a topological invariant called the first
Chern number (see the article by Joseph Avron, Daniel Osadchy, and
Ruedi Seiler, PHYSICS TODAY, August 2003, page 38). For a generally
interacting system, the topological properties of the QH state can
be described by an effective topological field theory based on the
Chern¨CSimons theory.\cite{r14} Although Duncan Haldane constructed
a model of the QH effect without the external magnetic field, that
state still breaks TR symmetry.

For a long time it was widely believed that both TR symmetry
breaking and two-dimensionality are necessary for an insulator to be
topological, but in 2001 the first model of a TR-invariant
topological insulator was introduced.\cite{r15} That model was
originally defined in 4D, but TR-invariant topological insulators in
3D and 2D can be obtained through a simple dimension-reduction
procedure.\cite{r12} Shuichi Murakami, Naoto Nagaosa, and Zhang, and
in parallel MacDonald and colleagues at the University of Texas in
Austin, developed the theory of the intrinsic spin Hall effect in
doped semiconductors and identified spin-orbit coupling as the
crucial ingredient; later, Murakami, Nagaosa, and Zhang extended the
theory to TR-invariant insulators. Kane and Mele first introduced
the topological band theory of TR-invariant QSH insulators in 2D and
showed that they fall into two distinct topological classes,
generally referred to as the ${\rm Z_2}$ classification.\cite{r9}
That beautiful topological band theory was soon generalized to three
dimensions.\cite{r11} The two of us and our colleagues have
developed a unifying topological field theory that defines the
general concept of a topological insulator in terms of a physically
measurable topological field theory.\cite{r12}

We now have two precise definitions of TR-invariant topological
insulators, one in terms of noninteracting topological band theory11
and one in terms of topological field theory.\cite{r12} If we
approximate an insulator with noninteracting electrons filling a
certain number of bands, the topological band theory can evaluate an
explicit topological invariant that can give only binary values of 0
or 1: a ${\rm Z_2}$ classification that defines trivial and
nontrivial insulators. For materials with inversion symmetry, a
powerful algorithm developed by Fu and Kane\cite{r4} can be easily
integrated into electronic structure calculations to numerically
evaluate the topological band invariant. However, since all
insulators in nature are necessarily interacting, it is important to
have a general definition of topological insulators that is valid
for interacting systems and is experimentally measurable. Both
problems were solved with the topological field theory,\cite{r12}
which can be generally defined for all insulators, with or without
interactions. In the noninteracting case, both definitions agree.
Surprisingly, the topological field theory can be explained in terms
of elementary concepts in undergraduate-level electromagnetism.

Inside an insulator, the electric field ${\bf E}$ and the magnetic
field ${\bf B}$ are both well defined. In a Lagrangian-based field
theory, the insulator's electromagnetic response can be described by
the effective action $S_0=1/8\pi\int d^3xdt(\epsilon E^2-1/\mu
B^2)$, with $\epsilon$ the electric permittivity and $\mu$ the
magnetic permeability, from which Maxwell's equations can be
derived. The integrand depends on geometry, though, so it is not
topological. To see that dependence, one can write the action in
terms of $F_{\mu\nu}$, the 4D electromagnetic field tensor:
$S_0=1/16\pi \int d^3xdtF_{\mu\nu}F^{\mu\nu}$. The implied summation
over the repeated indices $\mu$ and $\nu$ depends on the metric
tensor---that is, on geometry. (Indeed, it is that dependence that
leads to the gravitational lensing of light.) There is, however,
another possible term in the action of the electromagnetic field:
\begin{eqnarray}
S_\theta&=&\frac{\theta\alpha}{4\pi^2}\int d^3xdt{\bf E\cdot
B}\equiv\frac{\theta\alpha}{32\pi^2} \int d^3xdt
\epsilon_{\mu\nu\rho\tau}F^{\mu\nu}F^{\rho\tau}\nonumber\\
&=&\frac{\theta}{2\pi}\frac{\alpha}{4\pi}\int d^3xdt
\partial^\mu (\epsilon_{\mu\nu\rho\tau} A^\nu\partial^\rho
A^\tau) \label{CS}
\end{eqnarray}
where $\alpha=e^2/\hbar c\simeq 1/137$ is the fine-structure
constant, $\theta$ is a parameter, and $\epsilon_{\mu\nu\rho\tau}$
is the fully asymmetric 4D Levi-Civita tensor. Unlike the Maxwell
action, $S_\theta$ is a topological term---it depends only on the
topology of the underlying space, not on the geometry. Written using
the field tensor, the term is independent of the metric.

Since the ${\bf E}$ field is invariant under TR, whereas the ${\bf
B}$ field changes sign, $S_\theta$ naively breaks TR symmetry. For a
periodic system, however, there are two values of $\theta$, namely
$\theta=0$ or $\theta=\pi$, that preserve the TR symmetry.\cite{r12}
One can easily understand that conclusion by an analogy. If we have
a 1D ring with a magnetic flux inside, a general value of the flux
$\Phi$ would break the TR symmetry. However, for two special values
of the flux, $\Phi=0$ or $\Phi= hc/2e$, an electron's wavefunction
changes its phase by 0 or $\pi$ when the electron circles the ring
either clockwise or counterclockwise, and TR symmetry is maintained.

If we integrate out all the microscopic fermionic degrees of freedom
to obtain the effective action $S_\theta$, all nonmagnetic
insulators in the universe would fall into two distinct topological
classes, described by effective topological field theories with
$\theta=0$ or with $\theta=\pi$. Unlike $\epsilon$ and $\mu$, the
physically measurable $\theta$ parameter is universally quantized,
with the two possible values defining the topologically trivial and
nontrivial insulators, respectively---the ${\rm Z_2}$ classification
again.

Such classification is valid for a periodic system. For a real solid
with a finite boundary, a topological insulator is insulating only
in the bulk; it has an odd number of gapless Dirac cones on the
surface that describe conducting surface states. If we uniformly
cover the surface with a thin ferromagnetic film, an insulating gap
also opens up on the boundary; the TR symmetry is preserved in the
bulk but broken on the surface. The last identity of the equation
above for $S_\theta$ shows that the bulk topological term is in fact
a total derivative, expressible as a surface term, given by the
expression in parenthesis. That surface term is the same
Chern-Simons term that describes the topological field theory of the
QH state. In the QH field theory, the term's coefficient specifies
the value of the Hall conductance. \cite{r14} Here the coefficient
of $\theta=\pi$ translates into a Hall conductance of $1/2 e^2/h$,
half the conductance of the first QH plateau. That value is uniquely
associated with the single Dirac cone on the surface of topological
insulators. Any random disorder can change a system's Hall
conductance only by an integer multiple of $e^2/h$, thus the half-QH
conductance of $1/2 e^2/h$ can never be reduced to zero by
disorder---the surface states are topologically robust.

\section{Outlook}

\begin{figure}[htpb]
    \begin{center}
        \includegraphics[width=3.5in]{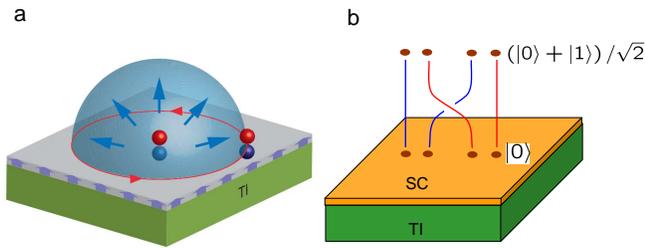}
    \end{center}
    \caption{Novel behavior is predicted for topological insulators. (a) When a topological
insulator (TI, green) is coated by a thin ferromagnetic layer
(gray), each electron (red sphere) in the vicinity of the surface
induces an image monopole (blue sphere) right beneath it.\cite{r12}
When one electron winds around another (red circle), it will
experience the magnetic flux (arrows in the blue dome) carried by
the image monopole of the other, so that the electron-monopole
composite, called a dyon, obeys fractional statistics. (b) When a TI
is coated by an s-wave superconductor (SC), the superconducting
vortices are Majorana fermions---they are their own antiparticles.
Exchanging or braiding Majorana vortices, as sketched here, leads to
non-Abelian statistics.\cite{r17} Such behavior could form the basis
for topological quantum computing.}
    \label{fig5}
\end{figure}

The field of topological insulators is growing rapidly, and many
remarkable experiments have been carried out. In nonlocal transport
measurements in a series of HgTe devices, the W${\rm
\ddot{u}}$rzburg group confirmed that transport current is carried
by the QSH edge states. The topological insulators ${\rm Bi_2Te_3}$
and ${\rm Bi_2Se_3}$ have been fabricated in nanoribbon form at
Stanford and by molecular-beam epitaxy at Tsinghua University.
Scanning tunneling microscopy experiments have been carried out at
Princeton, Stanford and Tsinghua universities to probe the
topological surface states. Preliminary transport measurements
indicate dominant contribution from surface states.

Solving Maxwell's equations with the topological term included leads
to predictions of novel physical properties characterized by exotic
excitations. The 2D QSH insulator is predicted to have fractional
charge at the edge and spin-charge separation in the bulk. In
introductory physics classes we learned that a point charge above a
metal or an insulator can be viewed as inducing an image charge
below the surface. A point charge above the surface of a 3D
topological insulator is predicted to induce not only an image
electric charge but also an image magnetic monopole below the
surface,\cite{r12} as shown in figure 5a. Such a composite object of
electric and magnetic charges, called a dyon, would obey neither
Bose nor Fermi statistics but would behave like a so-called anyon
with any possible statistics. Dislocations inside a 3D topological
insulator contain electronic states that behave similarly to QSH
edge states.

Axions are weakly interacting particles postulated to solve some
puzzles in the standard model of particle physics\cite{r16} (see the
article by Karl van Bibber and Leslie Rosenberg, PHYSICS TODAY,
August 2006, page 30). Those elusive particles are also predicted to
exist inside topological magnetic insulators, systems for which the
$\theta$ parameter above becomes dependent on position and time.
Majorana fermions are distinct from the familiar Dirac fermions:
They are their own antiparticles. There is still no conclusive
evidence for Majorana fermions in nature. But when a superconductor
is close to the surface of a topological insulator, Majorana
fermions are predicted to occur inside vortices (see figure
5b).\cite{r17}

Besides teaching us about the quantum world, the exotic particles in
topological insulators could find novel uses. For example, image
monopoles could be used to write magnetic memory by purely electric
means, and the Majorana fermions could be used for topological
quantum computing.\cite{r18}

Albert Einstein insisted that all fundamental laws of physics should
be expressed in terms of geometry, and he exemplified that ancient
Greek ideal by formulating the theory of gravity in terms of the
geometrical curvature of space and time. Physicists are now pursuing
Einstein's dream one step further, exploring the fundamental laws
expressed in terms of topological field theory. The standard model
of elementary particles contains a topological term that is
identical to the $S_\theta$ term that defines topological
insulators. Even if only a small number of the predicted exotic
particles are observed in topological insulators, our fundamental
understanding of nature would be greatly enhanced. Such tabletop
experiments could become a window into the standard model\cite{r16}
and help reveal the alluring beauty and mysteries of our universe.

\bibliography{ref_final}

\begin{widetext}
\appendix
\section{Models of topological insulators}
The essence of the quantum spin Hall effect in real materials can be
captured in explicit models that are particularly simple to solve.
The two-dimensional topological insulator mercury telluride can be
described by an effective Hamiltonian that is essentially a Taylor
expansion in the wave vector ${\bf k}$ of the interactions between
the lowest conduction band and the highest valence band:\cite{r2}
\begin{eqnarray}
H(k)&=&\epsilon({\bf k})\mathbb{I}+\left(\begin{array}{cccc}M({\bf k})&A(k_x+ik_y)&0&0\\A(k_x-ik_y)&-M({\bf k})&0&0\\
0&0&M({\bf k})&-A(k_x-ik_y)\\0&0&-A(k_x+ik_y)&-M({\bf k})\end{array}\right)\nonumber\\
\epsilon({\bf k})&=&C+D{\bf k}^2,~M({\bf k})=M-B{\bf k}^2\label{BHZ}
\end{eqnarray}
where the upper $2\times 2$ block describes spin-up electrons in the
s-like E1 conduction and the p-like H1 valence bands, and the lower
block describes the spin-down electrons in those bands. The term
$\epsilon({\bf k})\mathbb{I}$ is an unimportant bending of all the
bands ($\mathbb{I}$ is the identity matrix). The energy gap between
the bands is $2M$, and $B$, typically negative, describes the
curvature of the bands; $A$ incorporates interband coupling to
lowest order. For $M/B < 0$, the eigenstates of the model describe a
trivial insulator. But for thick quantum wells, the bands are
inverted, $M$ becomes negative, and the solution yields the edge
states of a quantum spin Hall insulator. Another model\cite{r9} for
the 2D topological insulator with a honeycomb lattice can also be
simply solved to gain explicit understanding.

The 3D topological insulator in the ${\rm Bi_2Te_3}$ family can be
described by a similar model:\cite{r6}
\begin{eqnarray}
H(k)&=&\epsilon({\bf k})\mathbb{I}+\left(\begin{array}{cccc}M({\bf k})&A_2(k_x+ik_y)&0&A_1 k_z\\A_2(k_x-ik_y)&-M({\bf k})&A_1 k_z&0\\
0&A_1 k_z&M({\bf k})&-A_2(k_x-ik_y)\\A_1 k_z&0&-A_2(k_x+ik_y)&-M({\bf k})\end{array}\right)\nonumber\\
\epsilon({\bf k})&=&C+D_1k_z^2+D_2k_\perp^2,~M({\bf
k})=M-B_1k_z^2-B_2k_\perp^2\label{Zhangetal}
\end{eqnarray}
in the basis of the Bi and Te bonding and antibonding pz orbitals
with both spins. The curvature parameters $B_1$ and $B_2$ have the
same sign. As in the 2D model, the solution for $M/B_1 < 0$
describes a trivial insulator, but for $M/B_1 > 0$, the bands are
inverted and the system is a topological insulator.

%\end{minipage}

\end{widetext}

\end{document}